\documentclass[prd,nofootinbib,preprint,superscriptaddress]{revtex4-1}
\usepackage{comment}
\usepackage{amsmath, amssymb, amsthm, graphicx, epsfig, fancyhdr,epsfig}
\usepackage{physics}
\usepackage{braket}
\usepackage{tikz-feynman}
\tikzfeynmanset{compat=1.1.0}
\usepackage{hyperref}
\usepackage{tikzsymbols}

\usepackage{float}
\usepackage{xcolor}

\newcommand{\be}{\begin{equation}}
\newcommand{\ee}{\end{equation}}
\newcommand{\bea}{\begin{eqnarray}}
\newcommand{\eea}{\end{eqnarray}}

\newcommand{\ag}[1]{\textcolor{blue}{[Anish: #1]}}

\begin{document}
\title{Irreducible Gravitational Wave Background as a Particle Detector}

\author{Anish Ghoshal}
\email{A.Ghoshal@sussex.ac.uk}
\affiliation{Department of Physics and Astronomy, University of Sussex,
Brighton, BN1 9RH, United Kingdom}

\author{Angus Spalding}
\email{angus.spalding1@gmail.com}
\affiliation{School of Physics and Astronomy, University of Southampton,
Southampton SO17 1BJ, United Kingdom}

\author{Graham White}
\email{G.A.White@soton.ac.uk}
\affiliation{School of Physics and Astronomy, University of Southampton,
Southampton SO17 1BJ, United Kingdom}

\begin{abstract}
Beyond the standard model particles can imprint a temporary early matter-dominated epoch on a stochastic gravitational-wave background. For a freely propagating background whose production has ceased, we show that the resulting feature contains two characteristic frequencies that determine the particle decay rate $\Gamma$ and the abundance-weighted mass $Y_iM$, independently of the microphysical origin of the gravitational-wave source. Using numerical scans, we derive semi-analytic relations between these characteristic frequencies and the BSM particle parameters, and quantify the parameter ranges over which these relations remain valid. For backgrounds that remain actively sourced, the end of early matter domination continues to define a source-independent cosmological frequency scale controlled by $\Gamma$, although the second feature, and hence the inference of $Y_iM$, is source dependent. Once a particle model and production mechanism are specified, the measured combinations can be translated into masses and microscopic couplings. Remarkably, decay lengths targeted by upcoming long-lived-particle searches map directly onto the nanohertz range in which NANOGrav has reported evidence for a stochastic gravitational-wave background.

\end{abstract}

\maketitle

\tableofcontents

\section{Introduction}
Stochastic gravitational-wave (GW) backgrounds provide us a unique opportunity to probe the early Universe, as they carries crucial information from pre-BBN (Big Bang Nucleosynthesis) epochs that are otherwise inaccessible  \cite{Caprini_2018, Bin_truy_2012, Roshan2024, van_Remortel_2023, Christensen_2018}. Consequently, the study of gravitational-wave backgrounds has become a major focus of current research, with next generation experiments expected to probe 12 decades of frequencies with sufficient GW strain sensitivity to shed light on the conditions of the first second of the Universe \cite{LIGOScientific:2016aoc, LIGOScientific:2016sjg, LIGOScientific:2017bnn, LIGOScientific:2017vox, LIGOScientific:2017ycc, LIGOScientific:2017vwq, Badurina:2021rgt, Graham:2016plp, Graham:2017pmn, Badurina:2019hst, Punturo:2010zz, Hild:2010id, LIGOScientific:2016wof, Reitze:2019iox, Baker:2019nia, AEDGE:2019nxb, Sesana:2019vho, Garcia-Bellido:2021zgu, Carilli:2004nx, Janssen:2014dka, Weltman:2018zrl, EPTA:2015qep, EPTA:2015gke, NANOGRAV:2018hou, Aggarwal:2018mgp, NANOGrav:2020bcs}. The LIGO–Virgo collaboration \cite{LIGOScientific:2016aoc, LIGOScientific:2016sjg} detected gravitational waves from black hole mergers a decade ago, while pulsar timing array (PTA) collaborations have very recently reported strong evidence for a stochastic GW background at nano-Hertz frequency \cite{Carilli:2004nx, Janssen:2014dka, Weltman:2018zrl, EPTA:2015qep, EPTA:2015gke, NANOGrav:2023gor, NANOGrav:2023hvm}. \\
These results have stimulated a wide range of beyond–Standard Model possibilities for cosmological GW production in the early Universe. The majority of these sources are transient, meaning they are no longer being sourced in our context, in particular primordial tensor modes from inflation \cite{Caprini_2016,Caprini_2018,Caldwell:2018giq,Boyle_2008,Boyle:2005se,Barman:2023ktz}, first-order phase transitions \cite{Caprini_2016, Hindmarsh:2015qta}, and annihilating domain walls \cite{Vilenkin:1981zs, Vachaspati:1984gt, Blanco_Pillado_2017, Hiramatsu_2010}. This is unlike for instance Domain Walls before annihilation \cite{Roshan:2026yon} and cosmic string gravitational wave sources \cite{Battye:2026whd}. \\ 
In the standard cosmological history after reheating, the Universe is radiation dominated until relatively late times, implying that gravitational-wave energy densities redshift with same rate as the dominant background, so it remains constant in time. However, any deviation from standard radiation domination modifies this behaviour~\cite{Pearce_2024,BARENBOIM2016430,Assadullahi_2009,Alabidi_2013,Kohri_2018,Inomata_2019, Ferreira:2025zeu, Ferreira:2026uzi, Spalding:2026pmp}. A temporary change in the equation of state alters the relative red-shifting between gravitational waves and the background energy density, imprinting a characteristic distortion on the observed GW spectrum ~\cite{Seto:2003kc, Boyle:2005se, Boyle:2007zx, Kuroyanagi:2008ye, Nakayama:2009ce, Kuroyanagi:2013ns, Jinno:2013xqa, Saikawa:2018rcs, Chen:2024roo, Iania:2026mtp}. \\
In this regard, we consider a period of early matter domination induced by heavy metastable particles \cite{Chung_1999, Datta:2025vyu, Ghoshal:2025iil,McDonald:1989jd,Moroi:1999zb, Visinelli:2009kt,Erickcek:2015jza,  Nelson:2018via,Cheek:2023fht,Cirelli:2018iax,Gouttenoire:2019rtn,Allahverdi:2020bys, Allahverdi:2021grt,Allahverdi:2022zqr}. Such long-lived particles (LLPs) are well motivated in many extensions of the Standard Model, including models of neutrinos \cite{Datta:2025vyu, Murayama:2025thw, Borboruah:2025hai}, axions \cite{Cheek:2025gvx}, supersymmetry \cite{Spalding:2026pmp}, extra dimensions \cite{Ghoshal:2025dmi}, and dark matter \cite{Cheek:2025gvx, Datta:2025vyu}, and are the focus of a rapidly expanding experimental program \cite{Gonzalez_Suarez_2021,jeanty2025longlivedparticlestheoryexperimental, Lee_2019,Alimena_2020, Verhaaren, Okada:2026evy}. Dedicated searches, including MATHUSLA, DUNE, and SHiP, are expected to significantly extend the sensitivity of these searches in the coming years \cite{DUNE:2015lol,Berryman:2019dme,Batell:2020vqn,Curtin:2018mvb,Bernardi:1985ny,Gorbunov:2021ccu,SHiP:2015vad}. Massive particle species with sufficiently small decay rates naturally redshift slower than radiation and can temporarily dominate the energy density of the Universe before decaying back into the thermal bath sourcing a period of early matter domination. \\
Such a deviation from standard radiation domination imprints two characteristic features in the GW background spectrum, corresponding to the onset and termination of the non-standard period \cite{Berbig:2023yyy, Borboruah:2024eha, Borboruah:2024eal, Bernal:2020ywq, Datta:2022tab, Datta:2023vbs, Chianese:2024nyw, Ghoshal:2022ruy, Datta:2025yow, Borboruah:2025hai, Cheek:2025gvx}. We show that these features in the GW spectrum encode the mass, decay width, and initial abundance of the long-lived particles. In particular, the two characteristic frequencies map directly to the decay rate $\Gamma$ and the product $M Y_i$. Consequently, within a given particle physics model, measuring these frequencies enables the underlying Lagrangian parameters, such as the particle mass and couplings governing the decay rate, to be inferred.\footnote{We should note that the decay of Long-Lived Particles can themselves generate a gravitational wave background either via graviton Bremmstrahlung \cite{Nakayama:2018ptw,Ghoshal:2022kqp,Choi_2024, Murayama:2025thw} or via decay to gravitons \cite{Strumia:2025dfn,Nakayama:2025xkn} however these are in the frequency range several orders larger than those considered in this work.}  A schematic illustration of this inference chain from gravitational wave observables to particle physics parameters is shown in Fig.~\ref{fig:GWB_inference_chain}. Production mechanisms include, but are not limited to, thermal production with freeze-out \cite{Datta:2025vyu}, inflaton decay \cite{Chung_1999, Allahverdi:2002nb}, number changing scattering processes \cite{Bringmann:2021tjr}, the misalignment mechanism \cite{PRESKILL1983127, Abbott:1982af, Dine:1982ah, Turner:1983he}, and gravitational production during inflation \cite{Lebedev:2022cic, Feiteira:2025rpe}. \\
The imprint of early matter domination on gravitational-wave backgrounds from long-lived particles has previously been studied in specific right-handed-neutrino scenarios for inflationary backgrounds \cite{Ghoshal:2025dmi} and cosmic-string backgrounds \cite{Datta:2025vyu}. Here, we generalise this framework to arbitrary gravitational-wave backgrounds and to any BSM particle capable of generating an early matter-dominated epoch, while providing an updated and improved numerical analysis. This is shown in Table~\ref{tab:prior_work_comparison}.

\begin{table}[t]
\centering
\small
\begin{tabular}{p{0.27\linewidth}p{0.38\linewidth}p{0.25\linewidth}}
\hline
\textbf{Work} & \textbf{GWB source} & \textbf{BSM particle} \\
\hline

Ref.~\cite{Borboruah:2025hai}
& Inflation
& Right-Handed Neutrinos
\\

Ref.~\cite{Datta:2025vyu}
& Cosmic strings
& Right-Handed Neutrinos
\\

Present work
& Generic GWB
& Generic BSM Particle
\\

\hline
\end{tabular}
\caption{\textit{Comparison with closely related studies of early matter domination and gravitational-wave backgrounds. Previous works have considered specific BSM particles generating early matter domination and specific GWB sources; the present work generalises these scenarios and provides an updated analysis.}}
\label{tab:prior_work_comparison}
\end{table}
\textit{This paper is organised as follows:} In Section~\ref{sec:EMD}, we study early matter domination driven by BSM particles, derive the conditions under which it occurs, and track the evolution of the equation-of-state parameter numerically. In Section~\ref{sec:GWBs}, we show that the resulting gravitational-wave feature is characterised by two frequencies and obtain numerical semi-analytic relations connecting them to the underlying model parameters. Finally, we conclude in Section~\ref{sec:conclusion}. 

\section{Early Matter Domination from BSM Particles}
\label{sec:EMD}
In this section we analyse how a long-lived particle species can generate a temporary period of early matter domination. We first derive analytic expressions describing the scaling behaviour of the system, obtaining relations for the temperature at which the species begins to dominate the energy density, $T_{\rm dom}$, and the temperature at which its decay restores radiation domination, $T_{\rm end}$, in terms of the particle mass, initial abundance, and decay rate. We then numerically solve the Boltzmann equations governing the evolution of the matter and radiation energy densities.
\subsection{Analytic Approximations}
We consider a general BSM particle species $X$ of mass $M$ that is decoupled from the thermal bath while still relativistic and subsequently evolves independently of the radiation plasma. The comoving number density of $X$ is characterised by its initial yield $Y_i \equiv n_X/s$, where $s$ is the entropy density. 
When the species behaves as matter the corresponding energy densities of matter and radiation are
\begin{equation}
    \rho_X = M\,n_X = M\,Y_i\,s,
    \qquad
    \rho_R = \frac{\pi^2}{30}\,g_*\,T^4,
\end{equation}
with $s = (2\pi^2/45)\,g_*\,T^3$ assuming no entropy dumps occur. The ratio then evolves as
\begin{equation}
    \frac{\rho_X}{\rho_R}
    = 
    \frac{M\,Y_i\,s}{\rho_R}
    =
    \frac{4}{3}\,Y_i\,\frac{M}{T},
\end{equation}
which grows as the temperature decreases. Matter domination occurs when $\rho_X = \rho_R$, defining
\begin{equation}
    T_{\rm dom} \;\simeq\; \frac{4}{3}\,Y_i\,M.
    \label{eq:Tdom_general_Y}
\end{equation}
Having found the temperature at which the particle dominates the energy budget of the universe we now look to find when it stops dominating the universe returning to radiation domination. This calculation is much simpler as we simply match the decay rate to the Hubble parameter at the decay temperature
\begin{equation} \Gamma = H(T_{\mathrm{end}}) = \sqrt{\frac{4\pi^3 g_*}{45}}\, \frac{T_{\mathrm{end}}^2}{M_{\mathrm{Pl}}}, \end{equation}
Rearranging for $T_{\mathrm{end}}$ gives
\begin{equation}
T_{\mathrm{end}}
=
\left(\frac{45}{4\pi^3 g_*}\right)^{1/4}
\sqrt{\Gamma\, M_{\mathrm{Pl}}}\simeq 0.24 \sqrt{\Gamma\, M_{\mathrm{Pl}}}
\end{equation}
Having found the analytic approximations we now find the relations numerically.

\subsection{Numerical Results}
\label{sec:numeric}
\begin{figure}[h!]
\centering
\includegraphics[width=1\linewidth]{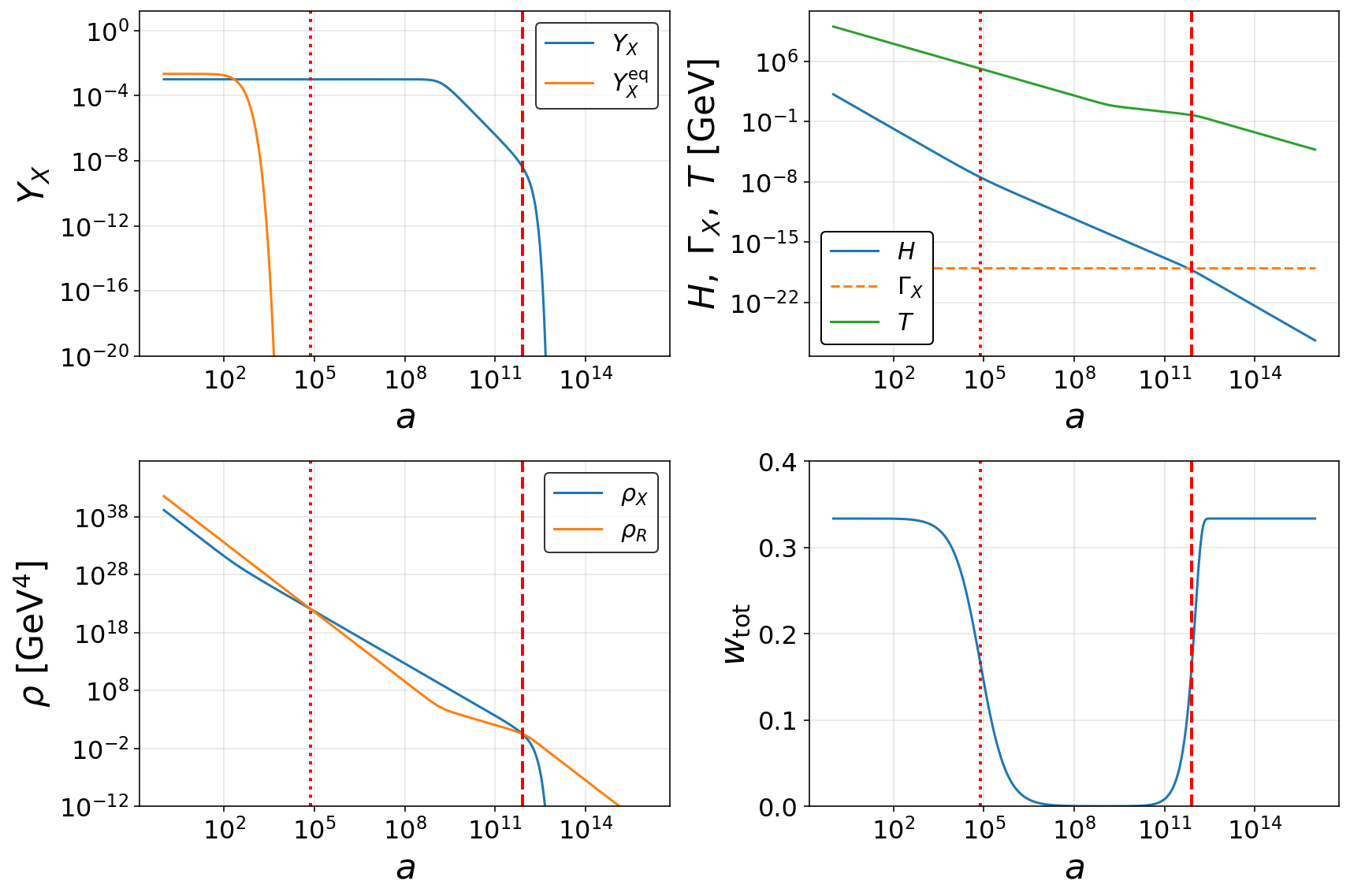} 
\caption{\it Benchmark Result for Early Matter domination by a heavy long-lived BSM particle. Plots show the evolution of the abundances (Top left), the Hubble parameter and temperature (Top Right), the physical energy densities (Bottom Left) and the equation of state parameter (Bottom Right). The dotted and dashed red vertical lines mark the beginning and end of the early matter-dominated era, respectively. Parameters are $M_X=10^{8}\ \text{GeV},\ Y_i=10^{-3}\ \Gamma=10^{-18}\text{GeV}$. We fixed $g_X=1$ however this is independent of the results. Discussion of results is given in the text.}
    \label{fig:Benchmark 1}
\end{figure}

The above analysis is useful to understand the scaling of the parameters and give analytical estimates however to determine accurately when matter domination begins ends and to determine the evolution of the equation of state we numerically solve the Boltzmann system for the energy densities for the thermal bath $\rho_R$ and the species X $\rho_X$. 
\begin{align}
    \dot \rho_X+3H(w_X+1)\rho_X=-\Gamma_X(\rho_X-\rho_X^{eq})\\
    \dot \rho_R+4H\rho_R=\Gamma_X(\rho_X-\rho_X^{eq})\ .
\end{align}
The Hubble parameter, $H$, is given by the Friedmann equation
\begin{equation}
    H(t) = \sqrt{\frac{8\pi}{3M_{\rm Pl}^2}\Big(\rho_X + \rho_R\Big)} .
\end{equation}
The equilibrium radiation energy density is determined by the temperature
\begin{equation}
\rho_R(T)=\frac{\pi^2}{30}g_*T^4 .
\end{equation}
The average energy per particle is calculated following maxwell-bolztmann statistics and is dependent on the mass of the particle and the temperature
\begin{equation}
    \langle E_X \rangle = M_X \frac{K_1\!\left(M_X/T\right)}{K_2\!\left(M_X/T\right)} + 3T,
\end{equation}
which allows us to calculate the equation of state parameter $w_X$,
\begin{equation}
    w_X=\frac{P_X}{\rho_X}=\frac{n_X T}{n_X \braket{E_X}}=\frac{T}{\braket{E_X}}\ .
\end{equation}
This value interpolates between radiation $w=1/3$ at $T\gg M$ and matter $w=0$ at $T\ll M$. Then the total equation of state parameter of the universe is given by,
\begin{equation}
    w_{tot}=\frac{w_X\rho_X+(1/3)\rho_R}{\rho_X+\rho_R}
\end{equation}
The equilibrium energy density similarly calculated in terms of the particle mass and the temperature, 
\begin{equation}
    \rho_X^{\rm eq}(T) \;=\; 
    \frac{g_X}{2\pi^2} \, M_X^2 \, T \, K_2\!\left(\frac{M_X}{T}\right)
    \left[ M_X \frac{K_1(M_X/T)}{K_2(M_X/T)} + 3T \right] .
\end{equation}
We provide a typical benchmark of the evolution of these variables in figure \ref{fig:Benchmark 1}.
\\
\\
Initially $T\gg M_X$ so the energy density redshifts with radiation. Then once $T\lesssim M$ $X$ behaves like matter and the energy density redshifts slower than radiation and comes to dominate the universe. Hubble parameter first shifts as radiation then after $T_{dom}$ shifts as matter domination then finally as Hubble drops below $\Gamma_X$ then the decays of X occur significantly. The abundance and physical energy densities then drop to a negligible amount and the Hubble redshifts like radiation again. The temperature drops until decays begin which can be seen as the point where $Y_X$ starts decreasing then radiation injection occurs slowing the decrease of the temperature. Then once decays complete the temperature shifts with expansion. The equation of state parameter begins with radiation domination at $w=1/3$ then as the particle acts like matter and comes to dominate the energy budget of the universe the equation of state parameter drops to zero, then when decays begin and the universe becomes radiation dominated the equation of state rises up to $w=1/3$ again, the red dotted line shows our analytical estimates. The estimate is good until decays back to radiation domination as we have assumed instantaneous decays for our analytic estimate. In what follows we shall use the numerical results at all time rather than the analytic estimates.\\
To find best fits to the domination and decay temperatures we scan over the parameters $Y,M,\Gamma$ and plot $T_{dom}$ against $(M,Y)$ and $T_{end}$ against $\Gamma$ to show the dependencies derived are correct and find best fits to the prefactors.
\begin{figure}[h!]
\centering
\includegraphics[width=1\linewidth]{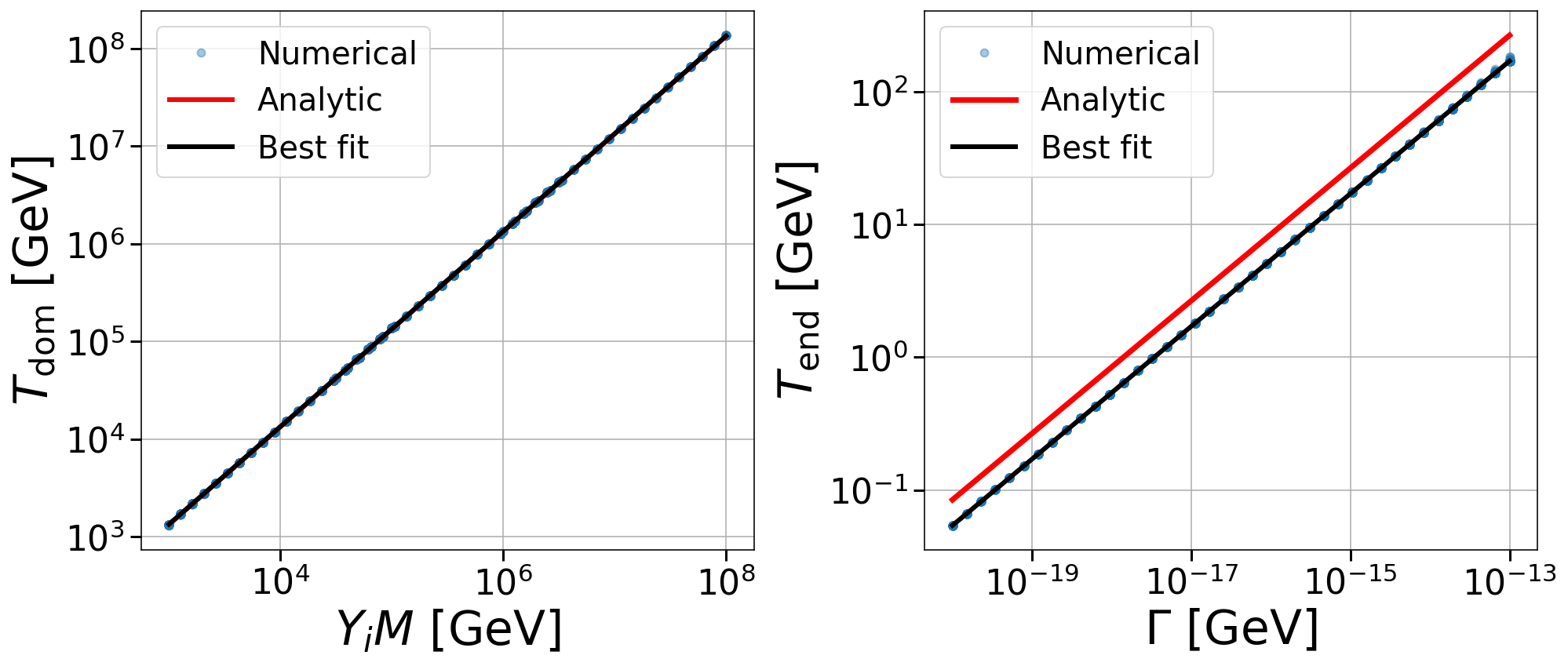} 
\caption{\it Scan for Early Matter domination by a heavy long-lived BSM particle. Parameters $(T_i=100\ M$ and scanned over the $\Gamma,M,Y_i$ parameter space. We fixed $g_X=1$ however this is larger independent of the results as is the initial temperature so long as the species begins as relativistic. As with the analytical estimate the domination temperature is determined by the product of mass and initial abundance, and the final temperature is determined by the decay rate. However with slightly reduced prefactors. Further discussion is given in the text.}
    \label{fig:MDscan}
\end{figure}
The best--fit relations extracted from the numerical scan shown in Fig.~\ref{fig:MDscan} are
\begin{equation}
    T_{\rm dom} = \frac{4}{3}\, Y_i M,\quad T_{\rm end} = 0.15\,\sqrt{\Gamma M_{\rm Pl}}
    \label{eq:tdomtdec}
\end{equation}

The corresponding logarithmic coefficients of determination coincide $R^2_{\rm log} = 0.999996$ demonstrating excellent agreement between the numerical results and the expected analytic scaling. The discrepancies between the numerical results and the analytic estimates arise from the simplifying assumptions used in the analytic derivations. In estimating the end of the matter-dominated era, we assume that the decay occurs instantaneously when $\Gamma_X=H$. In reality the decay proceeds continuously over a finite interval, so radiation domination is restored only after further evolution. To ensure that $X$ decays only after it has come to dominate the energy density, we require $T_{\rm end}<T_{\rm dom}$. This gives
\begin{equation}
    \frac{\Gamma M_{\rm Pl}}{Y_i^2M^2}
    <
    \left(\frac{4}{3\times0.15}\right)^2
    \simeq 79 .
    \label{eq:y_condition_general}
\end{equation}
This expression makes explicit the dependence on both the initial abundance
$Y_i$ and the particle mass $M$, and therefore applies to both thermal and
non-thermal production scenarios. To illustrate the resulting constraint,
we consider two examples:
\begin{enumerate}
    \item For thermally produced right-handed neutrinos,
    \begin{equation}
        \Gamma_N=\frac{\widetilde{m}M^2}{8\pi v^2},
    \end{equation}
    where $\widetilde{m}$ is the effective neutrino mass. Taking
    $Y_i=3.9\times10^{-3}$ and $v=174~{\rm GeV}$, the dependence on $M$
    cancels and Eq.~\eqref{eq:y_condition_general} becomes
    \begin{equation}
        \widetilde{m}
        <
        7.5\times10^{-8}~{\rm eV}.
    \end{equation}
    Although the cancellation of the heavy-neutrino mass is elegant, this
    condition requires a very small effective neutrino mass in the minimal
    model.

    \item Consider instead a massive scalar with
    $M_\phi=10^{12}~{\rm GeV}$ and $Y_i=10^{-1}$, decaying with
    \begin{equation}
        \Gamma_\phi=\frac{y^2}{8\pi}M_\phi .
    \end{equation}
    Equation~\eqref{eq:y_condition_general} then gives
    \begin{equation}
        y<1.3\times10^{-3}.
    \end{equation}
    This coupling is not especially small and is comparable to several
    Standard Model Yukawa couplings.
\end{enumerate}

The right-handed-neutrino example therefore requires a strongly suppressed
effective neutrino mass, whereas the scalar example can satisfy the
condition with a comparatively ordinary coupling. Indeed, five of the nine
charged-fermion Yukawa couplings in the Standard Model lie below the scalar
benchmark bound. These examples demonstrate that an early matter-dominated
era is a generic possibility for sufficiently long-lived particles with a
non-negligible initial abundance, rather than a feature tied to a particular
model. We now examine how this departure from radiation domination is imprinted on
a gravitational-wave background.

\section{Gravitational Wave Backgrounds with Early Matter Domination}\label{sec:GWBs}
The quantity of interest is the gravitational wave background, defined as the fractional energy density in gravitational waves per logarithmic frequency interval,
\begin{equation}
    \Omega_{\rm GW}(a,f)
    \equiv
    \frac{1}{\rho_{\rm tot}(a)}
    \frac{d\rho_{\rm GW}(a)}{d\ln f}\ .
\end{equation}
where $\rho_{GW}$ is the energy density of the gravitational wave background and $f$ denotes the frequency. We first consider a GWB whose production has ceased before the EMD transfer is evaluated. The subsequent tensor evolution is homogeneous, so that the observed spectrum factorizes as
\begin{equation}
 \Omega_{\rm GW}^{\rm obs}(f)=\mathcal{T}_{\rm EMD}(f)\,\Omega_{\rm GW}^{\rm ref}(f).
 \label{eq:source_factorization}
\end{equation}
Here $\Omega_{\rm GW}^{\rm ref}$ is the spectrum propagated through a reference radiation-dominated history, and $\mathcal{T}_{\rm EMD}$ depends on the post-production expansion history rather than on the microphysical production mechanism. We assume that there is a background GWB that has already been produced and is no longer being sourced and is freely propagating so that the only change is the cosmological evolution due to the change of the equation of state of the universe. This scenario may arise in first-order phase transitions, domain walls \cite{Roshan:2026yon}, scalar induced \cite{Mierna:2025vlm} and inflationary gravitational wave backgrounds. \\
Each observed frequency corresponds to a particular gravitational-wave mode. This mode does not immediately behave as a radiation component after production. While its physical wavelength is larger than the Hubble radius, the tensor perturbation is effectively frozen and does not propagate as a wave. In this regime the usual radiation-like redshifting description is not applicable, and the gravitational-wave energy fraction is not well-defined. As the Universe expands, the Hubble radius grows relative to the wavelength, and the mode eventually enters the horizon when its physical scale becomes comparable to the Hubble scale. From this point onward the mode oscillates and propagates freely, redshifting like radiation. Therefore, the evolution of the gravitational-wave background begins only after horizon entry, and $\Omega_{GW}$ evolves only once the corresponding mode is inside the horizon.\\
Horizon entry occurs when the physical wavenumber becomes comparable to the Hubble scale. This transition from superhorizon to subhorizon evolution is defined by
\begin{equation}
    k=a_{ent}H(a_{ent}),\quad k=2\pi f_0a_0
\end{equation}
where the subscript ``ent'' indicates the horizon entry, $f_0$ is the frequency as measured today and $a_0$ is the scale factor measured today which we calculate from our Boltzmann code assuming that between the end of the Boltzmann evolution and today there are no non-negligible entropy dumps.
\begin{equation}
    a_0=a_f\frac{T_f}{T_0}\left(\frac{g_*(T_f)}{g_*(T_0)}\right)^{\frac{1}{3}}
\end{equation}
where $a_f$ is the final scale factor of the evolution of Boltzmann equations, $T_f$ and $T_0$ are the final temperature of Boltzmann's and the temperature today respectively, $g_*$ is the degrees of freedom for the corresponding times.\\
This equation implicitly defines the entry scale factor for each frequency. In other words, the function $a_{ent}(f)$ provides the mapping between frequency space and cosmic time: higher-frequency modes satisfy this condition at earlier times (smaller scale factor), while lower-frequency modes enter later.\\
From the equation of motion for the tensor modes, it follows that in the superhorizon regime the modes are frozen, and therefore do not feel the effects of an early matter-dominated phase. Once a mode is inside the horizon, the fluid equations governing the energy densities of the gravitational-wave background and the total energy content of the Universe are 
\begin{equation}
    \dot{\rho}_{GW}+4H\rho_{GW}=0,\quad \dot{\rho}_{tot}+3H(w(a)+1)\rho_{tot}=0
\end{equation}
Differentiating with respect to the e-fold variable $\ln a$, the gravitational-wave background $\Omega_{\mathrm{GW}}$ evolves as
\begin{equation}
    \frac{d\ln \Omega_{GW}}{d\ln a}=3w(a)-1
\end{equation}
We now integrate this equation to obtain the full evolution of the spectrum. Evolution starts at the horizon-entry scale factor $a_{ent}(f).$ We therefore integrate from this point to some later time after we return to radiation domination. Integrating both sides and exponentiation gives,
\begin{equation}
    \Omega_{GW}(a_f,f)= \Omega_{GW}^{RD}(f)Exp\left[\int_{ln\ a_{ent}(f)}^{ln\ a_f}(3w(a)-1)d \ln a\right]
    \label{eq:MDeffect}
\end{equation}
where the superscript RD refers to the background under pure radiation domination. The above shows that there will be no departure from standard evolution during radiation domination $w=1/3$. Then as we have a period of early matter domination where $w<1/3$ there will be a suppression of the background before then returning the standard evolution predicted by radiation domination. \\
The analysis above relies on the instantaneous horizon entry approximation, where the tensor amplitude is assumed to remain constant for $k<aH$ and to immediately transition to radiation-like redshifting once $k>aH$. This is the correct result if the crossing occurs during radiation domination; however, if there is a deviation from radiation domination the modes will have an associated factor \cite{Boyle_2008}. This factor is a well-known result dependent on the equation of state parameter evaluated at horizon entry.
\begin{equation}
C(w)
=
\frac{\Gamma^2\!\left(\alpha+\tfrac{1}{2}\right)}{\pi}
\left(\frac{2}{\alpha}\right)^{2\alpha},\quad \alpha=\frac{2}{1+3w} \label{eq:C}
\end{equation}
where $\alpha$ should be evaluated at horizon re-entry, $k=aH$ assuming that $w$ is roughly constant during the crossing.\footnote{$\Gamma$ in Eq.~\ref{eq:C} denotes the Gamma function, not the decay rate.} As the return to radiation domination occurs from decays this is an accurate approximation.\footnote{Our analysis is valid if $X$ decays solely into radiation-like particles. This could be visible or dark radiation.} This factor is therefore 1 for radiation domination and early matter domination is $9/16$ so is an $\mathcal{O}(1)$ correction to the amplitude. We must also account for the effect of entropy injection on the source. Due to reheating, a given mode is mapped to an earlier effective emission time, resulting in a shift in the observed frequencies.\footnote{Should the source be fixed a certain time in the past with respect to the present then this shift is not there, see \cite{DEramo:2019tit, Borboruah:2025hai,Chianese:2024gee}.} We distinguish the comoving entropy from the GW transfer function. Define
\begin{equation}
 \mathcal S_{\rm com}(a)\equiv s(a)a^3,
 \qquad
 \Delta\equiv\frac{\mathcal S_{\rm com,f}}{\mathcal S_{\rm com,i}}.
 \label{eq:entropy_increase_definition}
\end{equation}
If the reference source spectrum is tabulated as a function of a pre-decay frequency coordinate, entropy production induces the horizontal remapping $f\mapsto f\Delta^{1/3}$. The full result can then be written
\begin{equation}
 \Omega_{\rm GW}(a_f,f)=
 \Omega_{\rm GW}^{\rm RD}\!\left(f\Delta^{1/3}\right)
 \mathcal T_{\rm EMD}(f),
 \label{eq:full_transfer_with_entropy}
\end{equation}
where
\begin{equation}
 \mathcal T_{\rm EMD}(f)=C[w(a_{\rm ent})]
 \exp\!\left[\int_{\ln a_{\rm ent}(f)}^{\ln a_f}
 (3w(a)-1)d\ln a\right].
 \label{eq:Sfunction}
\end{equation}
We can see that $\mathcal{T}_{\rm EMD}$ encodes both the thermal history of the universe and the effects of the early matter-dominated era, and we examine this function in detail in the following sections.

\subsection{Results}
\begin{figure}[h!]
\centering
\includegraphics[width=0.7\linewidth]{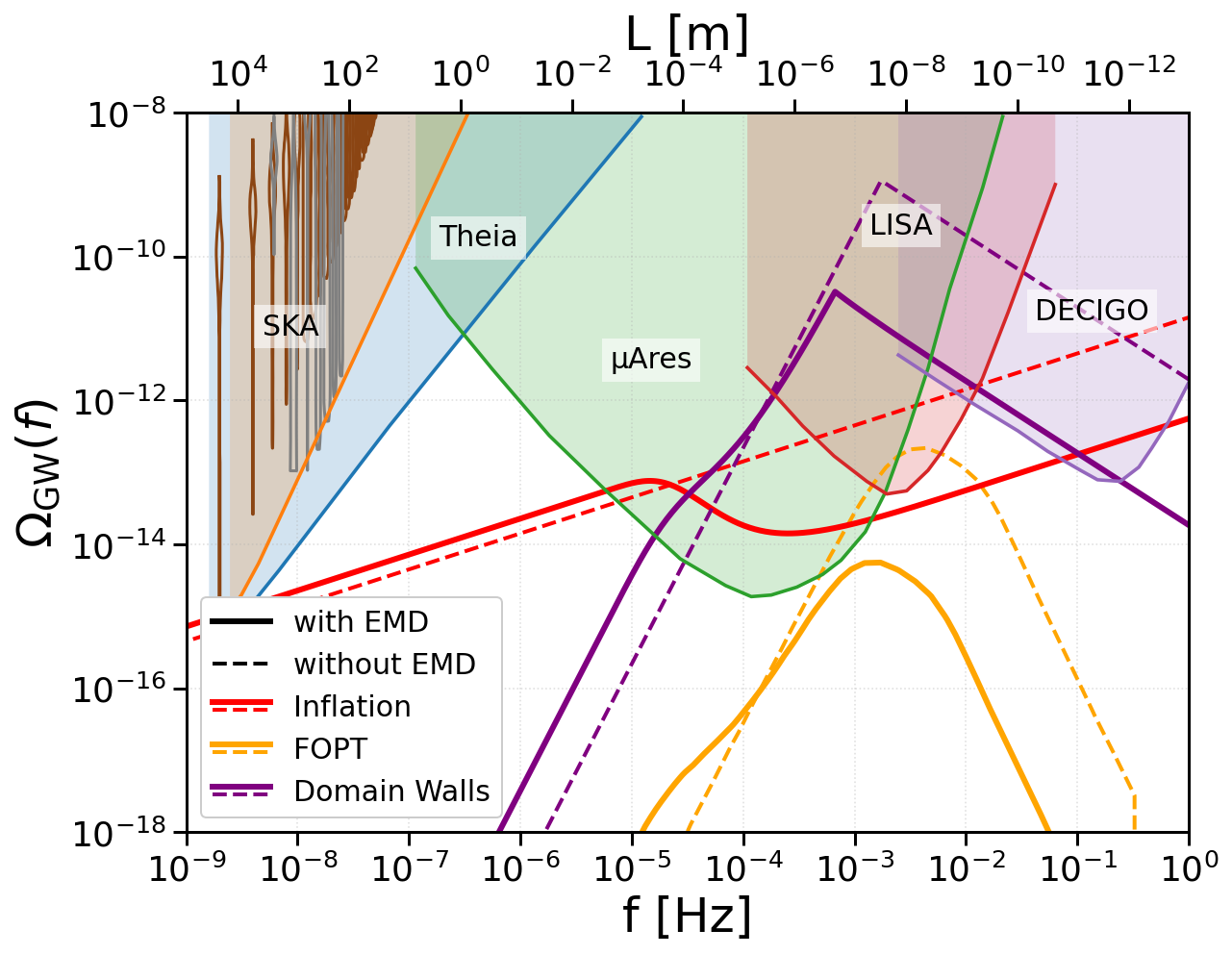} 
\caption{\it Benchmark GWBs with (solid lines) and without (dashed lines) an early matter-dominated era. Results are shown for inflationary (red), First-Order Phase Transitions (orange), and domain wall (purple). The shaded regions indicate the projected sensitivity of each detector, with the corresponding experiments labelled. Parameters for each benchmark shown in table \ref{tab:bench}. The upper x-axis indicates the location of the first feature for a particle with decay length L.}
    \label{fig:bench}
\end{figure}

\begin{figure}[t]
\centering
\includegraphics[width=1\linewidth]{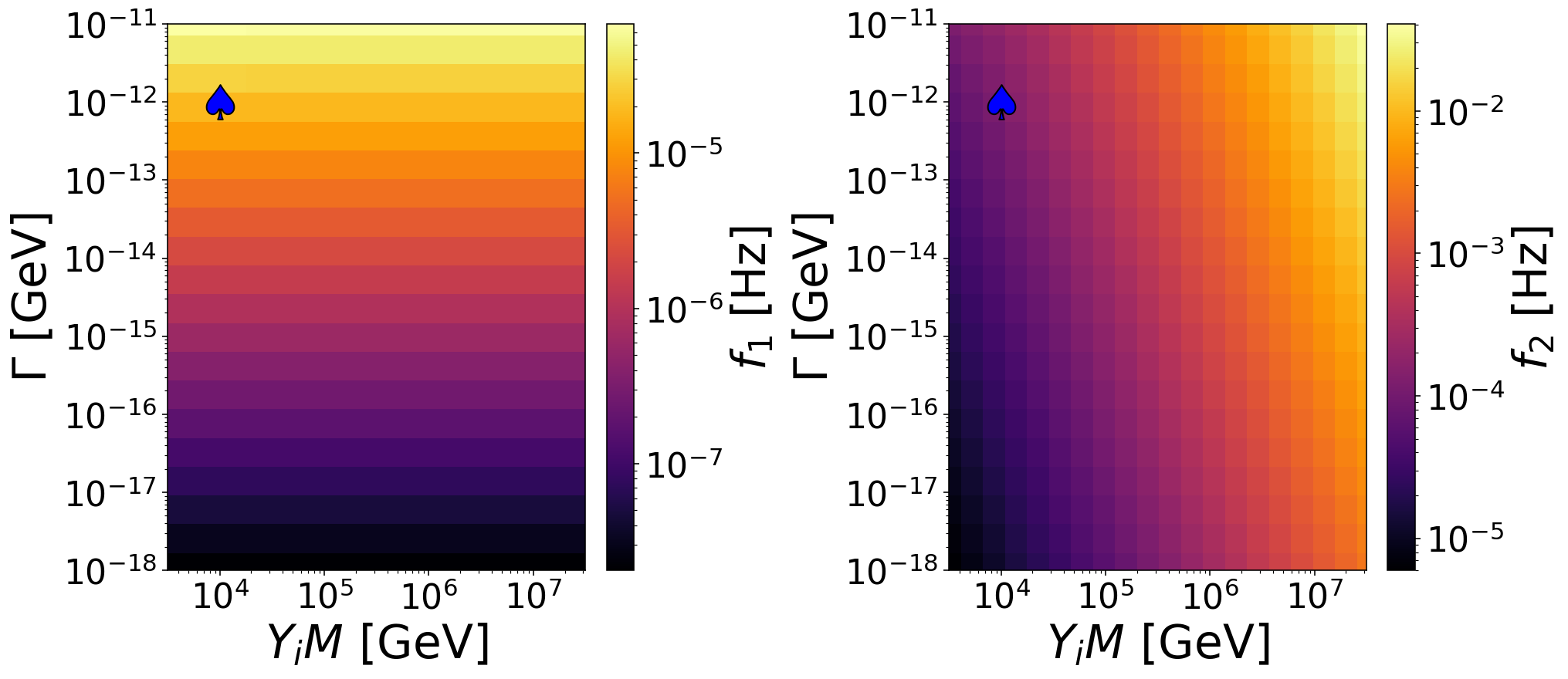} 
\caption{\it Scan of the characteristic frequencies over Decay rate and product of Mass and initial abundance recording the frequencies of the best fit parameters. We can see that $f_1$ is solely dependent on the decay rate corresponding to the end of early matter domination and that $f_2$ has a main dependence on $Y_iM$ and a mild dependence on the decay rate. The benchmark used in Fig \ref{fig:bench} is denoted by \textcolor{blue}{$\spadesuit$}. Further discussion given in the text.}
    \label{fig:fScan}
\end{figure}

To showcase the wide applicability of this framework, we present  Fig \ref{fig:bench} which displays gravitational-wave backgrounds from various sources with and without the subsequent period of early matter domination \footnote{For the benchmarks in Fig \ref{fig:bench} we assume the GWB is sourced during radiation domination, then a period of EMD occurs afterwards.}. As demonstrated by these benchmarks, gravitational wave observations can resolve both the onset and termination of an early period of matter domination for each source showing a clear and obvious feature compared to the background without a period of early matter domination. However, first-order phase transitions illustrate a complementary effect: in this case, the resulting signal can mask these features, allowing such scenarios to evade detection and thereby reopen regions of parameter space previously thought to be excluded.
The purpose of Fig.~\ref{fig:bench} is to demonstrate that the same post-production transfer function acts on distinct transient source shapes, not to rederive standard source spectra. The inflationary benchmark uses the tensor power-law spectrum and its radiation-era transfer as in Refs.~\cite{Ghoshal:2024gai,Planck:2018jri,Langlois:2010xc}. The annihilating-domain-wall benchmark uses the standard broken-power-law peak parametrization of Refs.~\cite{Kadota:2015dza,Hiramatsu:2013qaa}, neglecting plasma friction. The first-order phase-transition benchmark includes the sound-wave contribution evaluated with the numerical implementation of Ref.~\cite{Tian:2025zlo}, based on Ref.~\cite{Ellis:2020nnr}. Their parameters are listed in Table~\ref{tab:bench}. Since these expressions are standard and enter only through the reference spectrum $\Omega_{\rm GW}^{\rm ref}(f)$ in Eq.~\eqref{eq:source_factorization}, we do not reproduce their derivations here.
\begin{table}[h]
\centering
\footnotesize
\setlength{\tabcolsep}{1.5pt}
\renewcommand{\arraystretch}{0.80}

\begin{tabular}{|l l|}
\hline
\textbf{Source} & \textbf{Parameters} \\
\hline

Inflation 
& $\epsilon_{\rm inf} = 5\times 10^{-5},\; n_T = 0.5$ \\

FOPT 
& $\alpha = 0.25$, $\beta/H=50$, $T_n=1000$ GeV, $v_w=1$ \\

Domain & \\ walls 
& $\sigma_t = 2\times10^{17} \rm GeV,\; V_t = 6\times10^{28} GeV,\; \epsilon = 0.7,\; A = 0.8$\\
\hline
\hline

LLP& $\Gamma=10^{-12}$ GeV,\ $M=10^6$ GeV,\ $Y_i=0.01$,\ $MY_i=10^{4}$ \\
\hline
\end{tabular}

\caption{\it Benchmark parameters for each gravitational-wave source shown in Fig.~\ref{fig:bench}. The inflationary benchmark adopts a tensor power-law spectrum with the corresponding radiation-era transfer function, following Refs.~\cite{Ghoshal:2024gai,Planck:2018jri,Langlois:2010xc}. The annihilating-domain-wall benchmark is described using the standard broken-power-law parametrization of the peak spectrum from Refs.~\cite{Kadota:2015dza,Hiramatsu:2013qaa}, with plasma-friction effects neglected. For the first-order phase-transition benchmark, we include the sound-wave contribution using the numerical implementation of Ref.~\cite{Tian:2025zlo}, based on the treatment of Ref.~\cite{Ellis:2020nnr}. The specific production mechanism of the gravitational-wave background does not affect the particle-physics information extracted from the spectrum, since the two characteristic frequencies occur at the same locations for all freely propagating GWBs.}
\label{tab:bench}

\end{table}
\\
We now find the relationship between our initial parameters $(Y_i,M,\Gamma)$ and map to our transfer function as this is the key quantity that determines the GWB spectral features defined as,
\begin{equation}
    \mathcal T_{EMD}=C(w)Exp\!\left[
\int_{\ln a_{\rm ent}(f)}^{\ln a_f}
\left(3w(a)-1\right)d\ln a
\right]
\end{equation}
The constant radiation domination corresponds to $\mathcal T_{EMD}=1$. In the radiation-dominated regime at both early and late times, $w = \tfrac{1}{3}$, so the exponent vanishes and $\mathcal T_{EMD}$ remains constant, resulting in two flat plateaus. During an ideal matter-dominated interval, $w=0$, and Eq.~\eqref{eq:Sfunction} gives
\begin{equation}
    \mathcal T_{EMD}\simeq 0.6\ Exp\!\left[
\int_{\ln a_{\rm ent}(f)}^{\ln a_f}
\left(-1\right)d\ln a
\right]=\mathcal{O}(1)\frac{a_{ent}}{a_f}
\end{equation}
During early matter domination  we have, $k=a_{ent}H(a_{ent}),\ H\propto a^{-3/2}$ therefore we have, $a_{ent}\propto f^{-2}$ meaning $ \mathcal T_{EMD}\propto f^{-2}$ during early matter domination. To model this behaviour we fit the numerical results using the smooth broken power-law form\footnote{The validity of this fit is discussed in the Appendix material in section \ref{sec:validity}}
\begin{equation}
 \mathcal T_{\rm EMD}^{\rm fit}(f)=
 \frac{1+(f/f_2)^2}{1+(f/f_1)^2},
 \qquad f_1<f_2.
 \label{eq:Sfit}
\end{equation}
The three asymptotic regimes are
\begin{equation}
 \mathcal T_{\rm EMD}^{\rm fit}(f)\simeq
 \begin{cases}
 1, & f\ll f_1,\\[2pt]
 (f_1/f)^2, & f_1\ll f\ll f_2,\\[2pt]
 (f_1/f_2)^2, & f\gg f_2.
 \end{cases}
 \label{eq:Sfit_asymptotic_limits}
\end{equation}
Thus the intermediate branch has the expected $f^{-2}$ scaling, while the high-frequency modes form a fully suppressed plateau. The fitted $f_1$ and $f_2$ are template parameters extracted from the smooth spectrum; they are associated with, but need not be numerically identical to, instantaneous horizon scales evaluated at the density crossings.
The frequency \(f_1 (f_2)\) corresponds to modes entering the horizon near the end (start) of the early matter-dominated era. Modes with \(f<f_1\) enter after radiation domination resumes and therefore remain unaffected. Modes with \(f>f_2\) are already subhorizon before the deviation from radiation domination and thus experience the full suppression. These two frequencies are the GWB observables of interest as described in Fig \ref{fig:GWB_inference_chain}. These frequencies corresponding to the start and end of early matter domination therefore we will relate them to our initial parameters $MY_i$ and $\Gamma$.\\
We numerically solve the Boltzmann equations to obtain $w(a)$ for each case, then solve Eq.~\ref{eq:Sfunction} and fit to Eq.~\ref{eq:Sfit} to determine the dependence of the characteristic frequencies on the decay rate, initial abundance, and mass of the particle. The validity of these fits and more details are provided in section \ref{sec:validity}. The results of this scan are shown in Fig.~\ref{fig:fScan}. 
\\
We found $f_1$ is solely dependent the decay rate. By contrast $f_2$ has strong dependence on the product $Y_i M$, however when redshifting the frequency to the present there is a slight dependence on the length of early matter domination that occurred giving a slight dependence on the decay rate. This comes from the entropy dump at the end of the early matter domination period. The scan for $f_1$ yields the best-fit relation
\begin{equation}
    f_1 = 20.6\,\left(\frac{\Gamma}{\rm GeV}\right)^{1/2}\ \text{Hz}
    \label{eq:f1}
\end{equation}
while the scan for $f_2$ gives the best-fit relation
\begin{equation}
f_2 = 2.77\times10^{-5}\,\left(\frac{Y_i M}{\rm GeV}\right)^{2/3}
\left(\frac{\Gamma}{\rm GeV}\right)^{1/6} \ \text{Hz}\ .
\label{eq:f2}
\end{equation}
Once a model has been specified the Lagrangian parameters can be determined by specifying the initial abundance and writing the decay rate in terms of the mass and Lagrangian couplings. These best-fit expressions are not universally valid. The fit for $f_1$ remains accurate up to the bound given in Eq.~\ref{eq:y_condition_general} and can be used in future analysis. In contrast, the fit for $f_2$ is only reliable for decay rates at least two orders of magnitude below this bound. This can be understood from the fact that the best-fit result assumes a fully matter-dominated phase with $w = 0$. However, for short periods of early matter domination, such as those arising for decay rates close to the boundary of Eq.~\ref{eq:y_condition_general}, the equation of state never truly reaches $w = 0$. Achieving this would require $\rho_X \gg \rho_R$, whereas in this regime one typically has only $\rho_X > \rho_R$. As a result, the system remains in a partially matter-dominated phase, and the idealised scaling underlying the best fit is no longer valid. In the Appendix material (Sec. \ref{sec:appendix}), we derive this scaling, finding agreement up to a differing overall prefactor.\\
To make the complementarity with collider searches for LLPs more transparent, we rewrite the characteristic frequency $f_1$ in terms of the proper decay length, $L=1/\Gamma$, in metres, 
\begin{equation}
    f_1 = 2.9\times 10^{-7}\left(\frac{1\ \mathrm{m}}{L}\right)^{1/2}\ \text{Hz}\ \label{eq:f1lab}.
\end{equation}
This is shown in Fig \ref{fig:bench} as the top x-axes. As a representative example, particles with decay lengths of order $L \sim \mathcal{O}(100)\,\mathrm{m}$ which is accessible in MATHUSLA \cite{Feng:2017vli,Curtin:2018mvb} corresponds to a frequency $f_1 \simeq 2\times 10^{-8}\,\text{Hz}$, placing the signal squarely within the sensitivity range of pulsar timing array experiments, \cite{EPTA:2015qep, EPTA:2015gke, NANOGrav:2023gor, NANOGrav:2023hvm}, as well as future facilities such as SKA \cite{Janssen:2014dka,Weltman:2018zrl,Carilli:2004nx} and Theia \cite{Garcia-Bellido:2021zgu}. \footnote{For earlier estimates in the context of right-handed neutrinos, see \cite{Borboruah:2025hai}.}\\
Remarkably, this frequency lies in the region where a stochastic gravitational-wave background has recently been reported by NANOGrav, suggesting that LLP scenarios with decay lengths accessible at colliders could leave observable imprints in the nano-hertz gravitational-wave spectrum. It is worth emphasizing that any parameter space satisfying Eq. \ref{eq:y_condition_general} will produce a feature at the frequency given by Eq. \ref{eq:f1lab}.
Importantly, this means that Eq.~\eqref{eq:f1lab} can generally be used. To show what decay lengths are then probed by each future experiment we provide the table \ref{tab:decay_lengths}.
 \begin{table}[h!]
\centering
\begin{tabular}{lcc}
\hline
\textbf{Detector} & \textbf{Frequency $f_1$ [Hz]} & \textbf{Decay Length $L$ [m]} \\
\hline
Theia / SKA & $10^{-9} - 10^{-7}$ & $10^{5} - 10^{1}$ \\
$\mu$Ares / LISA & $10^{-6} - 10^{-2}$ & $10^{-1} - 10^{-9}$ \\
DECIGO & $10^{-2} - 1$ & $10^{-9} - 10^{-13}$ \\
ET / CE & $1 - 10^{2}$ & $10^{-13} - 10^{-17}$ \\
\hline
\end{tabular}
\caption{\it Mapping between gravitational wave frequency $f_1$ and decay length $L$ using $L=1\,{\rm m}\left[(2.9\times10^{-7}\,{\rm Hz})/f_1\right]^2$. Lower frequencies correspond to larger decay lengths, while higher-frequency detectors probe increasingly microscopic scales.}
\label{tab:decay_lengths}
\end{table}
We can see that due to the large range of frequencies being probed by upcoming experiments a large range of particle decay lengths are being probed.\\
Once a specific particle physics model is specified, these frequencies can be mapped onto Lagrangian parameters, such as the particle mass and coupling constants through the decay rate. The general process of deducing the Lagrangian parameters from the gravitational wave observables are shown in figure \ref{fig:GWB_inference_chain}. \footnote{A recent follow-up to an earlier version of this work \cite{Spalding:2026pmp} showed that, for gravitinos and more generally for particles whose decay is gravitationally suppressed, the first break frequency $f_1$ maps directly onto the mass of the BSM particle. This illustrates that some classes of BSM particles can encode more detailed information in the gravitational-wave spectrum than others.}
\begin{figure}[H]
\centering
\resizebox{0.95\linewidth}{!}{%
\begin{tikzpicture}[
    node distance=3.0cm and 2.6cm,
    box/.style={
        rectangle, rounded corners, draw=black, fill=blue!5, thick,
        text centered, text width=4.0cm, minimum height=1.8cm,
        font=\bfseries\LARGE
    },
    arrow/.style={thick, -{Latex[length=3.5mm,width=2mm]}},
    eqnlabel/.style={midway, above, sloped, font=\Large\itshape},
    eqnlabelV/.style={pos=0.5, left, font=\Large\itshape}
]

\node[box] (gwb) {GWB\\ observables};

\node[box, above right=1.6cm and 3cm of gwb] (f2) {$f_{1}$};
\node[box, below right=1.6cm and 3cm of gwb] (f1) {$f_{2}$};

\node[box, right=3.5cm of f1] (ym) {$YM$};
\node[box, right=3.5cm of f2] (gamma) {$\Gamma$};

\node[box, right=3.5cm of ym] (mass) {$M$};
\node[box, right=3.5cm of gamma] (lag) {Lagrangian\\ coupling};

\draw[arrow] (gwb) -- (f2);
\draw[arrow] (gwb) -- (f1);

\draw[arrow] (f1) -- node[eqnlabel] {Eq.~\ref{eq:f2}$+\Gamma$} (ym);
\draw[arrow] (f2) -- node[eqnlabel] {Eq.~\ref{eq:f1}} (gamma);

\draw[arrow] (ym) -- node[eqnlabel] {Model } (mass);
\draw[arrow] (gamma) -- node[eqnlabel] {Model+$M$} (lag);
\end{tikzpicture}%
}

\caption{\it Schematic showing how gravitational-wave-background observables can be mapped onto particle-physics parameters. The characteristic frequency $f_1$, associated with the end of early matter domination, determines the decay rate $\Gamma$ through Eq.~\ref{eq:f1}. For a freely propagating GWB, the characteristic frequency $f_2$, associated with the onset of early matter domination, determines the product $Y_iM$ through Eq.~\ref{eq:f2} once $\Gamma$ is known. Separating $Y_iM$ into the particle mass $M$ and initial abundance $Y_i$, and relating $\Gamma$ to an underlying Lagrangian coupling, require specification of the particle model and production mechanism.}
\label{fig:GWB_inference_chain}
\end{figure}

\subsection{Non-Transient and Sourced GWBs}
We now consider gravitational-wave backgrounds that remain actively sourced
during the early matter-dominated era. In this case the gravitational-wave
fluid equation is modified by the addition of a source term,
\begin{equation}
    \dot{\rho}_{\rm GW}(t,f)
    +4H(t)\rho_{\rm GW}(t,f)
    =
    \mathcal{S}(t,f),
    \label{eq:sourced_gw_fluid}
\end{equation}
where \(\mathcal{S}(t,f)\) denotes the gravitational-wave energy injected per
unit time and per logarithmic frequency interval. Following the logic of the last section we obtain
\begin{equation}
    \frac{d\Omega_{\rm GW}(a,f)}{d\ln a}
    =
    \bigl(3w(a)-1\bigr)\Omega_{\rm GW}(a,f)
    +
    \frac{\mathcal{S}(a,f)}
         {H(a)\rho_{\rm tot}(a)}
    \label{eq:Omega_sourced_lna}
\end{equation}
Defining
\begin{equation}
    \mathcal{J}(a,f)
    \equiv
    \frac{\mathcal{S}(a,f)}
         {H(a)\rho_{\rm tot}(a)},
\end{equation}
the solution at a final scale factor \(a_f\) is
\begin{align}
    \Omega_{\rm GW}(a_f,f)
    &=
    \Omega_{\rm GW}(a_i,f)
    \exp\left[
        \int_{\ln a_i}^{\ln a_f}
        \bigl(3w(a)-1\bigr)d\ln a
    \right]
    \nonumber\\
    &\quad+
    \int_{\ln a_i}^{\ln a_f}
    d\ln a'\,
    \mathcal{J}(a',f)
    \exp\left[
        \int_{\ln a'}^{\ln a_f}
        \bigl(3w(a)-1\bigr)d\ln a
    \right].
    \label{eq:sourced_solution}
\end{align}
The first term is the freely propagating contribution considered previously,
while the second term contains gravitational waves continuously generated at
different times. Each contribution produced at \(a'\) is therefore multiplied
by the transfer factor
\begin{equation}
    \mathcal{T}(a',a_f)
    \equiv
    \exp\left[
        \int_{\ln a'}^{\ln a_f}
        \bigl(3w(a)-1\bigr)d\ln a
    \right].
    \label{eq:sourced_transfer}
\end{equation}

The lower-frequency feature \(f_1\) corresponds to the comoving Hubble scale
at the end of early matter domination,
\begin{equation}
    f_1
    =
    \frac{a_{\rm end}H_{\rm end}}{2\pi a_0}.
    \label{eq:f1_sourced}
\end{equation}
At \(a=a_{\rm end}\), the equation of state returns to radiation domination,
\(w=1/3\), and hence the dilution term in
Eq.~\eqref{eq:Omega_sourced_lna} vanishes. This transition occurs at the same
background time irrespective of whether the gravitational-wave background is
freely propagating or continuously sourced. The position of the feature is therefore unchanged:
\begin{equation}
    f_1^{\rm sourced}
    =
    f_1^{\rm transient}
    =
    \frac{a_{\rm end}H_{\rm end}}{2\pi a_0}\ .
    \label{eq:f1_unchanged}
\end{equation}
Consequently, the relation between \(f_1\) and the decay rate remains valid,
\begin{equation}
    f_1
    =
    20.6
    \left(
        \frac{\Gamma}{\mathrm{GeV}}
    \right)^{1/2}
    \mathrm{Hz}.
    \label{eq:f1_sourced_Gamma}
\end{equation}
We note, however, that source-dependent spectral structure may partially mask or mimic the EMD-induced feature, introducing a degeneracy in its observational identification even though its underlying cosmological scale remains fixed.

The situation is different for the second characteristic frequency. For a
transient background, \(f_2\) is associated with the comoving Hubble scale at
the onset of early matter domination,
\begin{equation}
    f_2^{\rm transient}
    =
    \frac{a_{\rm dom}H_{\rm dom}}{2\pi a_0}.
    \label{eq:f2_transient}
\end{equation}
All modes with \(f>f_2\) are already inside the horizon before matter
domination begins and therefore experience the complete suppression. For an
actively sourced background, however, gravitational waves at a fixed observed
frequency can be generated both before and during early matter domination.
Equation~\eqref{eq:sourced_solution} then gives
\begin{equation}
    \Omega_{\rm GW}^{\rm sourced}(a_f,f)
    =
    \int_{\ln a_i}^{\ln a_f}
    d\ln a'\,
    \mathcal{J}(a',f)\,
    \mathcal{T}(a',a_f),
    \label{eq:sourced_convolution}
\end{equation}
up to any pre-existing freely propagating contribution. The observed spectrum
is thus a convolution of the time-dependent source with the cosmological
transfer function. The location and shape of the feature associated with the
onset of matter domination consequently depend on the time and frequency
dependence of \(\mathcal{S}(a,f)\) and in general, $f_2^{\rm sourced}\neq f_2^{\rm transient}$. We therefore conclude that for a non-transient gravitational-wave background
the frequency \(f_1\) remains a source-independent probe of the end of early
matter domination and hence of the decay rate \(\Gamma\). By contrast, \(f_2\)
depends on the detailed evolution of the gravitational-wave source, so the
extraction of \(Y_iM\) becomes source dependent.
\section{Discussion \& Conclusion}\label{sec:conclusion}
A sufficiently long-lived particle can generate a temporary epoch of early matter domination (EMD), leaving a characteristic imprint on a stochastic gravitational-wave background. For a pre-existing, freely propagating GWB whose source has switched off, the effect of EMD factorizes into the reference spectrum and an EMD transfer function. The two characteristic break frequencies then determine the cosmological combinations
\begin{equation}
(f_1,f_2)\longrightarrow(\Gamma,Y_iM),
\end{equation}
through Eqs.~\eqref{eq:f1} and \eqref{eq:f2}. Recovering the individual particle parameters requires additional microscopic input: separating $Y_iM$ into the abundance $Y_i$ and mass $M$ requires a specified production history, while relating $\Gamma$ to a coupling depends on the particle model. Under these assumptions, the measured spectral features can be translated into particle masses and couplings, as illustrated by our right-handed-neutrino and scalar examples. The corresponding precision therefore inherits both the observational uncertainty in the reconstructed break frequencies and the theoretical uncertainty associated with the production mechanism.\\
For actively sourced backgrounds, the situation is less universal. The EMD epoch still defines the cosmological scales $f_{\rm end}$ and $f_{\rm dom}$, with $f_{\rm end}\propto\Gamma^{1/2}$ fixed by the end of matter domination. However, because the observed spectrum is determined by the convolution in Eq.~\eqref{eq:sourced_convolution}, the visibility and precise fitted location of these features can depend on the source history. In particular, the reconstruction of $Y_iM$ from the second scale is generally source dependent, whereas the decay-rate information associated with $f_{\rm end}$ remains considerably more robust. The EMD condition in Eq.~\eqref{eq:y_condition_general} also favours increasingly heavy particles, allowing gravitational-wave observations to probe masses far beyond those directly accessible at terrestrial experiments.\\
We considered decay rates relevant for future long-lived-particle searches at experiments such as DUNE \cite{DUNE:2015lol}, MATHUSLA \cite{Curtin:2018mvb}, and SHiP \cite{SHiP:2015vad}. Remarkably, we found that the corresponding decay lengths map directly onto the nanohertz frequency range in which NANOGrav has reported evidence for a stochastic gravitational-wave background. At the same time, gravitational-wave observations extend sensitivity to particle masses far beyond the reach of terrestrial colliders. Future detections of stochastic gravitational-wave backgrounds may therefore reveal otherwise inaccessible information about long-lived particles and non-standard epochs in the thermal history of the universe, establishing a direct complementarity between pulsar timing arrays, future gravitational-wave observatories, and laboratory searches for long-lived particles.\\
We are now entering an era in which gravitational-wave observatories can double as particle detectors.

\section*{Acknowledgements} 
GW and AS acknowledge the STFC Consolidated Grant ST/X000583/1. AS thanks the University of Southampton School of Physics and Astronomy for the support of a Mayflower PhD scholarship. A.G. acknowledges the support from the Royal Society, UK, Funding Reference: NIF\ R1\ 253963. Authors thank Germano Nardini, Pankaj Borah, Nimmala Narendra, Shuailiang Ge, Elliot Tole and Yuhsin Tsai for useful discussion and comments on the manuscript. Authors are indebted to Marek Lewicki for pointing out an inaccuracy in Eq. \ref{eq:full_transfer_with_entropy} in version 1, which consequently affected Fig. \ref{fig:bench}. The key results of this work Eq \ref{eq:f1} and Eq \ref{eq:f2} were not affected.

\appendix
\section{Breaking Frequency dependence}
\label{sec:appendix}
In this section, we derive analytic approximations for the two characteristic frequencies, which serve as our key observables. We recover the correct scaling with the input parameters $\Gamma, M, Y_i$, albeit with a slightly different numerical prefactor.\\
The first frequency $f_1$ from the plot is solely determined by the decay rate, we now provide a calculation showing this dependence. The frequency can be determined from the comoving horizon scale at the end of early matter domination redshifted to today,

\begin{equation}
f_1 =\frac{a_{\rm end}H_{\rm end}}{2\pi a_0} .
\end{equation}

Approximating $H_{\rm end}\simeq \Gamma$ and using entropy conservation,

\begin{equation}
\frac{a_{\rm end}}{a_0}
=
\frac{T_0}{T_{\rm end}}
\left(\frac{g_{*s,0}}{g_{*s}}\right)^{1/3}
\simeq 0.34\,\frac{T_0}{T_{\rm end}} ,
\end{equation}

where we used $g_{*s,0}=3.91$ and $g_{*s}=106.75$. Using the relation
$T_{\rm end} = 0.15\sqrt{\Gamma M_{\rm Pl}}$, this gives

\begin{equation}
f_1 =
\frac{T_0}{2\pi \cdot 0.15}
\left(\frac{g_{*s,0}}{g_{*s}}\right)^{1/3}
\frac{\Gamma^{1/2}}{\sqrt{M_{\rm Pl}}}
\simeq 36.0\,
\left(\frac{\Gamma}{\text{GeV}}\right)^{1/2}\,\text{Hz} .
\end{equation}

which is the scaling observed in the numerical scans, with the overall normalisation differing by a factor of $\mathcal{O}(1.5)$. This discrepancy arises from the simplifying assumption of an instantaneous decay, $H_{\rm end}\simeq \Gamma$.

Now we evaluate $f_2$, the frequency corresponding to the onset of early matter domination. Using the horizon-entry condition $k=aH$, the observed frequency today scales as

\begin{equation}
f(a_{\rm ent}) = \frac{k}{2\pi a_0}
= \frac{a_{\rm ent}H(a_{\rm ent})}{2\pi a_0} .
\end{equation}

It is convenient to consider the ratio of the two characteristic frequencies,

\begin{equation}
\frac{f_2}{f_1} = \frac{a_{\rm dom}H_{\rm dom}}{a_{\rm end}H_{\rm end}} .
\end{equation}

During the early matter-dominated era the Hubble rate scales as $H\propto a^{-3/2}$, giving

\begin{equation}
\frac{f_2}{f_1} = \left(\frac{a_{\rm end}}{a_{\rm dom}}\right)^{1/2}
= \left(\frac{\rho_{\rm dom}}{\rho_{\rm end}}\right)^{1/6} .
\end{equation}

At the onset of early matter domination $\rho_X=\rho_R$, so that

\begin{equation}
\rho_{\rm dom} = 2\rho_R = \frac{\pi^2}{15}g_* T_{\rm dom}^4 .
\end{equation}
At the end of early matter domination, using $H_{\rm end}\simeq \Gamma$ and the Friedmann equation,
\begin{equation}
\rho_{\rm end} = \frac{3M_{\rm Pl}^2\Gamma^2}{8\pi} .
\end{equation}
Combining these gives
\begin{equation}
\frac{f_2}{f_1}
=
\left(\frac{8\pi^3 g_*}{45}\right)^{1/6}
T_{\rm dom}^{2/3}
M_{\rm Pl}^{-1/3}
\Gamma^{-1/3} .
\end{equation}
Using $T_{\rm dom}=\frac{4}{3}Y_iM$, we obtain
\begin{equation}
f_2=f_1\left(\frac{8\pi^3 g_*}{45}\right)^{1/6}
\left(\frac{4}{3}Y_iM\right)^{2/3}
M_{\rm Pl}^{-1/3}
\Gamma^{-1/3} .
\end{equation}

Substituting the analytic expression for $f_1$ then gives

\begin{equation}
f_2 =5.5\times10^{-5}
\left(\frac{Y_iM}{\mathrm{GeV}}\right)^{2/3}
\left(\frac{\Gamma}{\mathrm{GeV}}\right)^{1/6}
\mathrm{Hz} .
\end{equation}
Once again we have derived the true scaling found by the Boltzmann evolution however with a slight different constant.
\section{Validity of Best-Fits}
\label{sec:validity}
\begin{figure}[h!]
\centering
\includegraphics[width=1\linewidth]{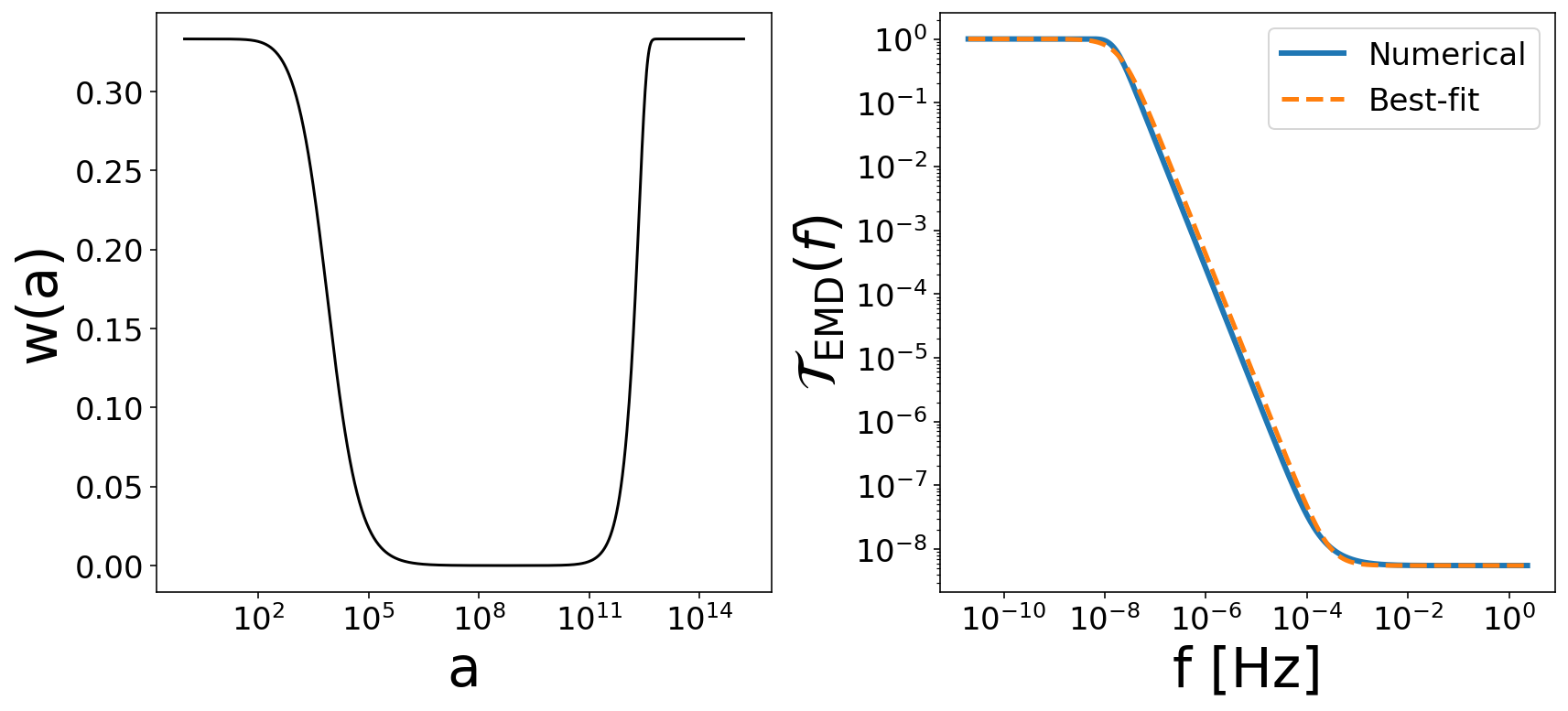}
\caption{\it Benchmark evolution of the dilution factor \(\mathcal T_{EMD}\) in Eq 30 of the main text for the initial parameters \(Y_i=10^{-2}\), \(M=10^6~\mathrm{GeV}\), and \(\Gamma=10^{-18}~\mathrm{GeV}\). \textbf{Left Panel:} Evolution of the equation-of-state parameter. \textbf{Right Panel:} Numerical evaluation of the dilution factor \(\mathcal T_{EMD}(f)\) as a function of frequency together with the corresponding best-fit function, demonstrating excellent agreement between the numerical result and the fit.}
\label{fig:Sbench}
\end{figure}

\begin{figure}[h!]
\centering
\includegraphics[width=0.48\linewidth]{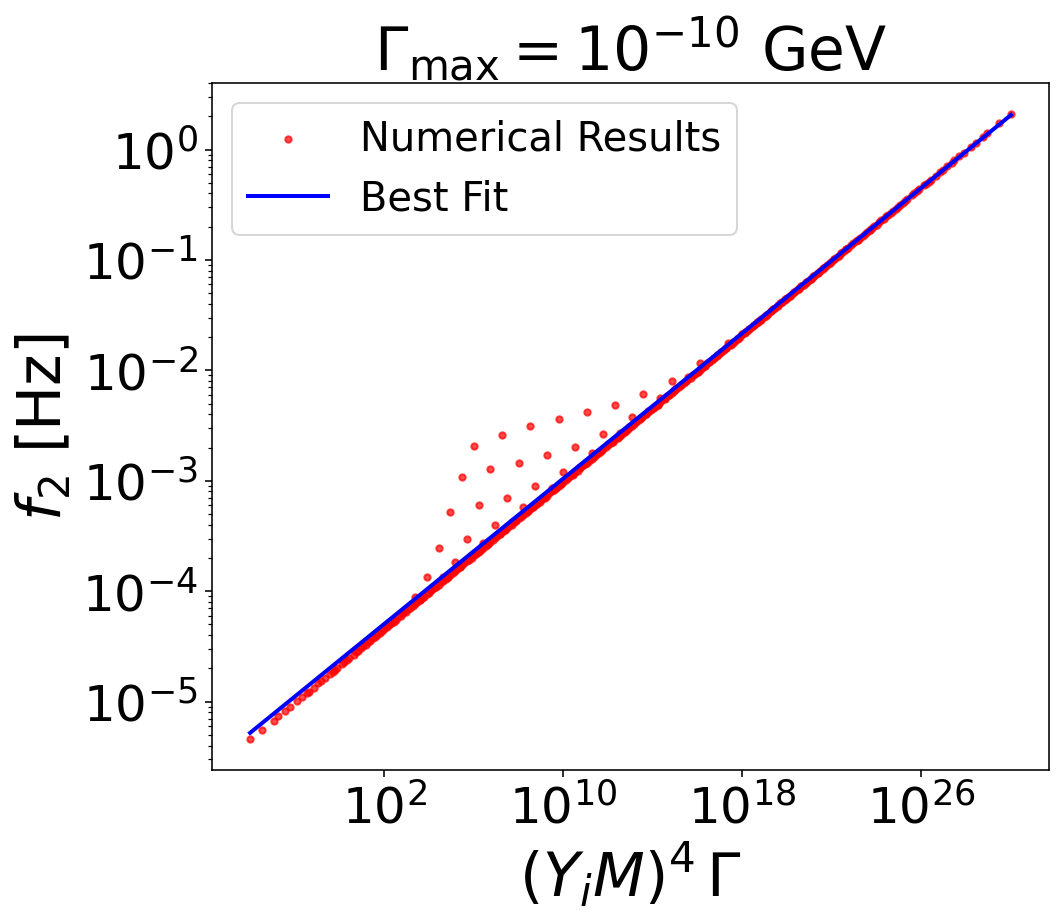}
\includegraphics[width=0.48\linewidth]{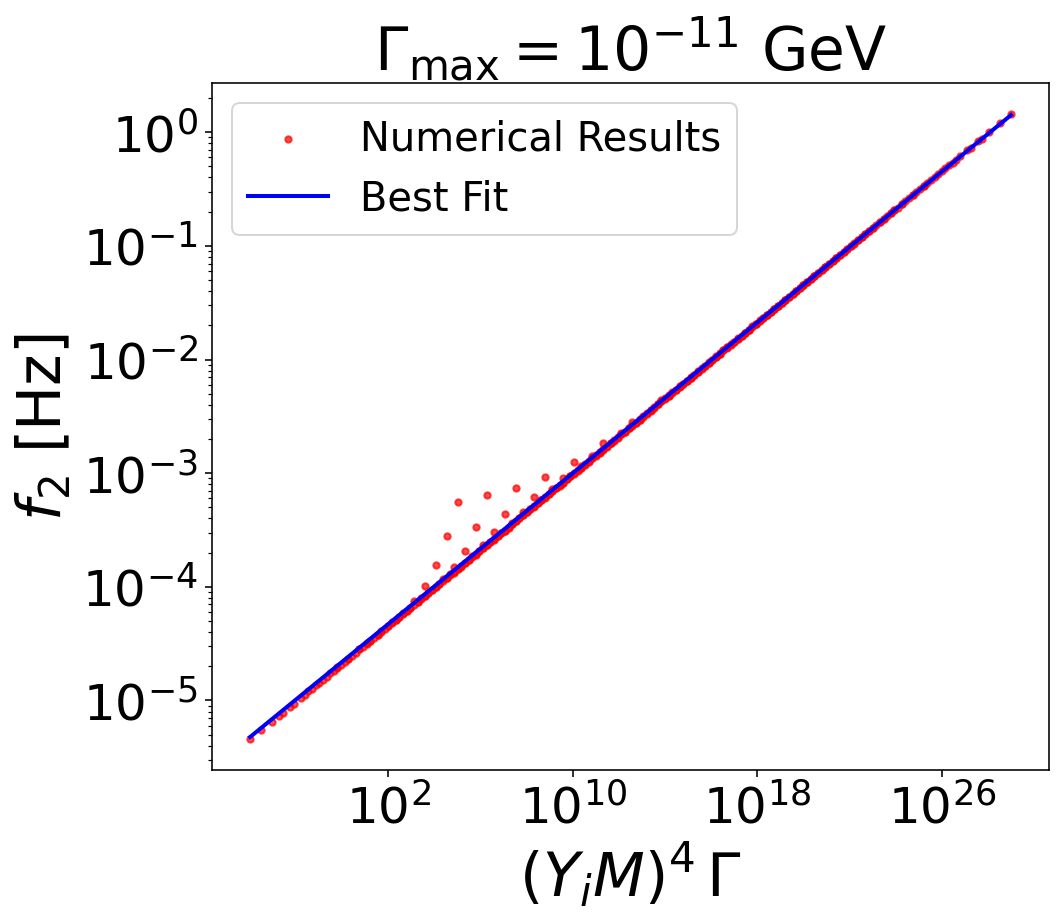}

\vspace{0.3cm}

\includegraphics[width=0.48\linewidth]{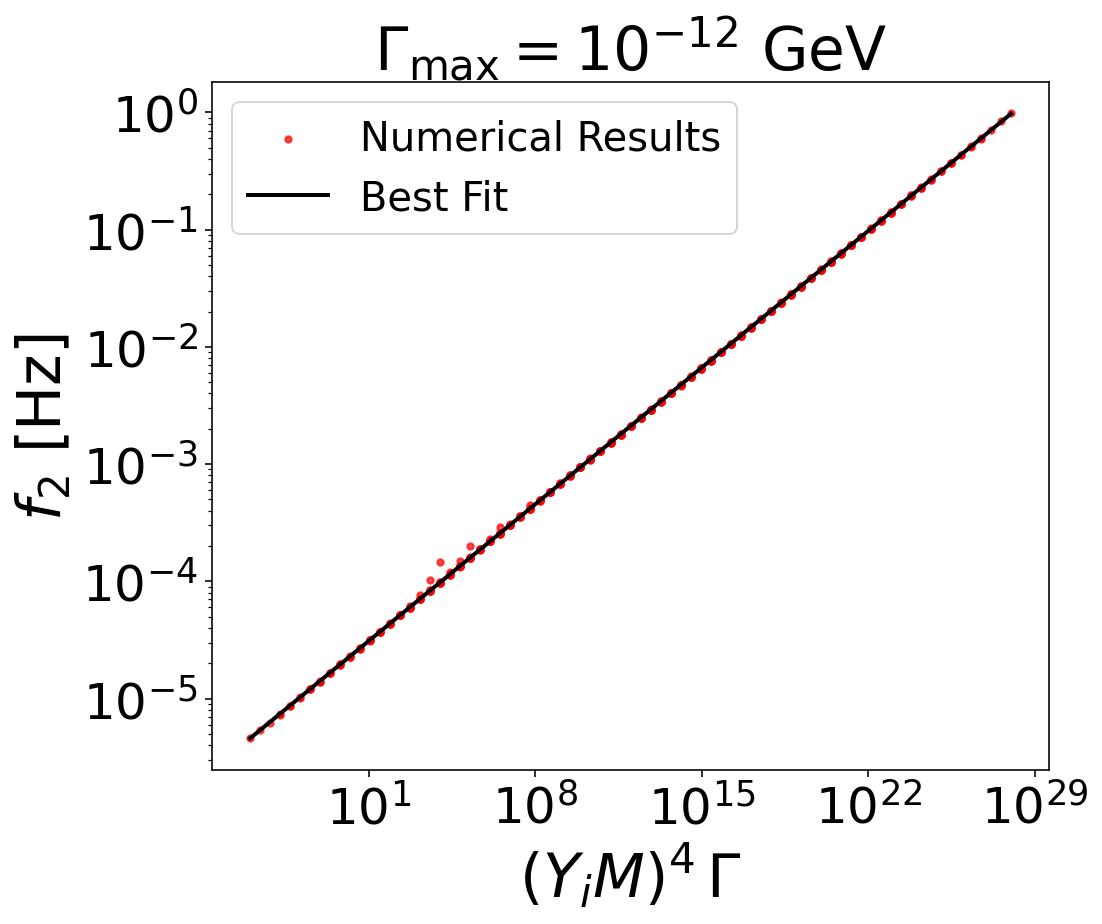}
\includegraphics[width=0.48\linewidth]{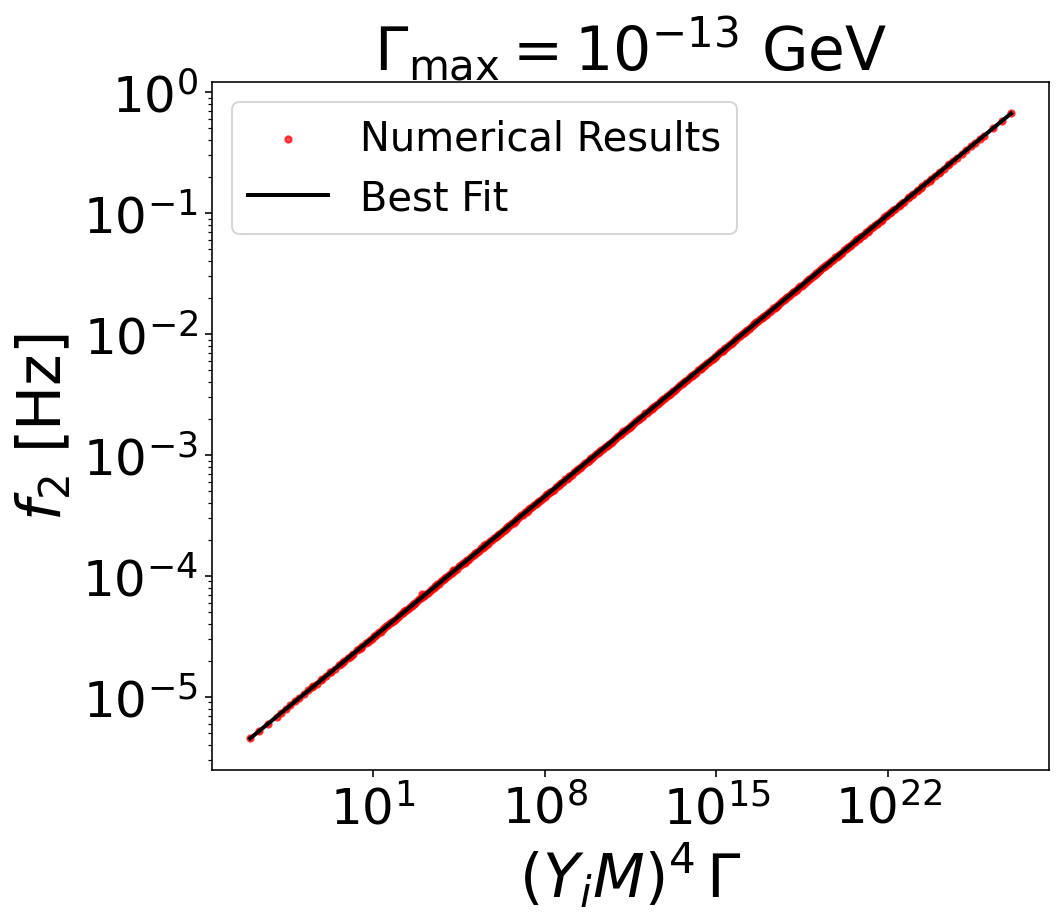}

\caption{\it Scan of $f_2$ for different maximum decay rates $\Gamma = 10^{-10}, 10^{-11}, 10^{-12}, 10^{-13}$ (ordered left-to-right, top-to-bottom). The maximum $\Gamma$ that gives early matter domination is approximately $\Gamma_{\rm max}=10^{-10}$ GeV. We can see that values of Gamma close to this bound do not lie on the best fit line, it is not until two orders of magnitude below that this can be seen as a good fit. }
\label{fig:f2GammaGrid}
\end{figure}

In this subsection we assess the validity of the best-fit relations given in Eqs. 35 and 36 of the main text. As a quantitative validation, we compute the logarithmic coefficient of determination for each benchmark, 
\begin{equation}
 R^2_{\log}=1-\frac{\sum_i[\log_{10}\mathcal T_i-\log_{10}\mathcal T_{{\rm fit},i}]^2}
 {\sum_i[\log_{10}\mathcal T_i-\langle\log_{10}\mathcal T\rangle]^2}.
 \label{eq:R2log_current}
\end{equation}
which measures the quality of the fit in log-space. We find $R^2_{\log} > 0.998$ throughout the scan, demonstrating that the template provides an excellent description of the numerical spectrum across the full frequency range. As a benchmark we provide Fig \ref{fig:Sbench}.
We now examine the validity of the extracted scaling relations for $f_1$ and $f_2$. Scanning over the parameter space, we find that the best-fit relation for $f_1$ remains robust provided the condition in Eq. 15 of the main text is satisfied. In contrast, the relation for $f_2$ in Eq. 36 of the main text breaks down when the duration of the early matter-dominated phase becomes too short. To illustrate this, we perform scans over increasing values of $\Gamma$, approaching the boundary of Eq. 15 of the main text, as shown in Fig.~\ref{fig:f2GammaGrid}. Near this boundary, the numerical values of $f_2$ deviate systematically from the best-fit scaling, and only recover the expected behaviour once $\Gamma$ is at least two orders of magnitude below the maximum value allowing early matter domination.

This behaviour has a clear physical origin. The derivation of the $f_2$ scaling assumes a well-defined matter-dominated phase with $w \simeq 0$. However, close to the boundary, the energy densities satisfy $\rho_M \sim \rho_R$, corresponding to an effective equation of state $w \simeq 1/6$, rather than $w=0$. As a result, the system never fully enters a matter-dominated regime, invalidating the assumptions underlying the analytic scaling. To show this we take various scans over Gamma up to approaching the boundary. 
We can see to get valid best fit the decay rate has to be less than roughly two order of magnitude than Eq. 15 of the main text however below that the best fit remains the same. For the $f_1$ best fit, one should be able to reproduce the fit by scanning from the boundary in Eq. 15 of the main text and to smaller decay rates. Therefore $f_1$ and as such the equation for decay length, Eq. 37 of the main text, are valid. This is due to the need for the best fit to achieve a $w=0$ which does not happen for early matter domination close to the boundary as this is $\rho_M\approx \rho_R$ corresponding to a $w\approx \frac{1}{6}$ not 0.

\section{Redshifting vs Entropy Dilution Discussion}

The suppression of a gravitational-wave background during a period of early matter domination (EMD) is often described in terms of entropy dilution. However, this suppression may equivalently be understood directly from the modified redshifting behaviour during the non-standard expansion history. In this subsection we clarify that these are not distinct physical effects, but rather two equivalent interpretations of the same underlying dynamics caused by the dominant component of the energy budget.\\
The observable used for gravitational wave backgrounds are 
\begin{equation}
    \Omega_{\rm GW}(a,f)
    \equiv
    \frac{1}{\rho_{\rm tot}(a)}
    \frac{d\rho_{\rm GW}(a)}{d\ln f}\ .
\end{equation}
where we have used the conventional notation in the literature, $\rho_{\rm tot}=\rho_C$, after the inflationary phase. In this context, the two quantities are equivalent. The effect of early matter domination is two-fold:
\begin{enumerate}
    \item Horizontal shift: Due to entropy injection increasing the temperature, modes undergo additional redshifting before reaching the present-day temperature.
    \item Vertical shift for some frequencies: Dilution due to the modified expansion history changing the evolution of $\rho_C$. This may be described in two equivalent ways:
    \begin{enumerate}
        \item The redshifting of $\rho_C$: $\rho_C$ redshifts slower than $\rho_{\rm GW}$, leading to a suppression of the overall spectrum.
        \item Entropy injection: After the EMD period has ended, the value of $\rho_{\rm tot}$ is larger than it would have been under uninterrupted radiation domination due to the entropy dump. This increases the background energy density relative to $\rho_{\rm GW}$, leading to a dilution of the gravitational-wave background.
    \end{enumerate}
\end{enumerate}
We now look to show these are equivalent. We begin with the interpretation used in our analysis. For modes experiencing the full suppression plateau, we take the large frequency limit of our best fit Eq. 34 of the main text
\begin{equation}
    \mathcal T_{EMD}(f\gg f_2)=\frac{f_1^2}{f_2^2},
\end{equation}
Using the scaling found for the characteristic frequencies,
\begin{equation}
    f_1 \propto \Gamma^{1/2},
    \qquad
    f_2 \propto
    (Y_i M)^{2/3}\Gamma^{1/6},
\end{equation}
gives
\begin{equation}
     \mathcal T_{EMD}^{\rm max}
    \propto
    \Gamma^{2/3}(Y_i M)^{-4/3}.
\end{equation}
Thus the suppression becomes larger for longer periods of matter domination, corresponding to smaller decay rates or larger initial abundances. The same result may alternatively be derived from the entropy dilution picture. During reheating, the decay of the dominant matter component injects entropy into the thermal bath. Defining the comoving entropy increase as \cite{Roshan:2026yon}
\begin{equation}
    \Delta
    \equiv
    \frac{S_f}{S_i}
    =
    \frac{s_f a_f^3}{s_i a_i^3}\simeq \frac{T_{\rm dom}}{T_{\rm end}}
\end{equation}
The fractional gravitational-wave energy density is diluted according to \cite{Roshan:2026yon}
\begin{equation}
    \Omega_{\rm GW}
    \rightarrow
    \Omega_{\rm GW}\Delta^{-4/3}.
\end{equation}
Using
\begin{equation}
    T_{\rm dom}\sim Y_i M,
    \qquad
    T_{\rm end}\sim \sqrt{\Gamma M_{\rm Pl}},
\end{equation}
we obtain
\begin{equation}
    \Delta^{-4/3}
    \sim
    \left(
    \frac{T_{\rm end}}{T_{\rm dom}}
    \right)^{4/3}
    \propto
    \Gamma^{2/3}(Y_i M)^{-4/3},
\end{equation}
which reproduces precisely the same scaling obtained from the redshifting derivation.

We therefore conclude that the suppression of the gravitational-wave background is fundamentally caused by the modified expansion history during EMD through its impact on $\rho_C$. This effect may be interpreted either as the differential redshifting between $\rho_{\rm GW}$ and the dominant matter component $\rho_C$, or equivalently as entropy dilution during reheating. These are simply two complementary descriptions of the same physical process.
\bibliographystyle{apsrev4-1}

\bibliography{ref}
\end{document}